%% file: main.tex
\documentclass{INTERSPEECH2023}

\interspeechcameraready 

\usepackage{tabularx}
\usepackage{xcolor}
\usepackage[mode=buildnew]{standalone}
\usepackage{multirow}
\usepackage{moresize}
\usepackage{scalerel}
\usepackage{circledsteps}

\title{E\MakeLowercase{ff}CRN: An  Efficient Convolutional Recurrent Network\\
for High-Performance Speech Enhancement}
\name{Marvin Sach$^\text{*}$ \quad 
Jan Franzen$^\text{*}$ \quad 
Bruno Defraene$^\circ$ \quad Kristoff Fluyt$^\circ$ \quad Maximilian Strake$^\text{*}$ \\ 
Wouter Tirry$^\circ$ \quad Tim Fingscheidt$^\text{*}$}
\address{
\begin{minipage}{0.45\textwidth}
\begin{center}
$^\text{*}$Technische Universität Braunschweig\\Institute for Communications Technology\\ 38106 Braunschweig, Germany
\end{center}
\end{minipage}
\begin{minipage}{0.45\textwidth}
\begin{center}
$^\circ$Goodix Technology (Belgium) BV\\ 3000 Leuven, Belgium
\end{center}
\end{minipage}
}
\email{ \{m.sach, j.franzen, m.strake, t.fingscheidt\}@tu-bs.de, \\ \{bdefraene, kfluyt, wtirry\}@goodix.com}

\begin{document}

\newcommand{\circled}[1]{\textcircled{\raisebox{-0.5pt}{\scriptsize #1}}}

\newcommand{\circledp}[1]{\raisebox{0.0pt}{\scriptsize {+}}\textcircled{\raisebox{-0.5pt}{\scriptsize #1}}}
\newcommand{\circledm}[1]{\raisebox{0.0pt}{\scriptsize {-}}\textcircled{\raisebox{-0.5pt}{\scriptsize #1}}}

\maketitle
 
\begin{abstract}
Fully convolutional recurrent neural networks (FCRNs) have shown state-of-the-art performance in single-channel speech enhancement. However, the number of parameters and the FLOPs/second of the original FCRN are restrictively high. 
A further important class of efficient networks is the CRUSE topology, serving as reference in our work. By applying a number of topological changes at once, we propose both an efficient FCRN (\texttt{FCRN15}), and a new family of efficient convolutional recurrent neural networks (\texttt{EffCRN23}, \texttt{EffCRN23lite}).
We show that our \texttt{FCRN15} (875K parameters) and \texttt{EffCRN23lite} (396K) outperform the already efficient \texttt{CRUSE5} (85M) and \texttt{CRUSE4} (7.2M) networks, respectively, w.r.t.\ PESQ, DNSMOS and $\Delta$SNR, while requiring about 94\% less parameters and about 20\% less \#FLOPs/frame. Thereby, according to these metrics, the FCRN/EffCRN class of networks provides new best-in-class network topologies for speech enhancement.
\end{abstract}
\noindent\textbf{Index Terms}: noise suppression, efficient networks, convolutional recurrent neural networks, speech enhancement

\section{Introduction}

In single channel noise suppression the aim is to estimate a clean speech signal from a noisy mixture of clean speech and interfering background noise. For real world applications, efficient methods enabling low-latency real-time processing and adhering to memory limitations are of utmost importance \cite{Fedorov2020, Braun2021, Tan2021, valinrnnoise}. Recently, neural networks have seen increasing use for this task with many of the prominent approaches estimating a complex spectral mask for the noisy speech in the short-time Fourier transform (STFT) domain \cite{Wang2018c, Strake2020b, Williamson2016, Hu2020b}.

Especially convolutional neural networks (CNNs) have been widely used for the task of speech enhancement \cite{Park2017, Strake2020, Xu2021a}. The sliding kernels allow for precise modelling of local dependencies in the speech spectra \cite{Strake2020}.

Recurrent processing has been employed to model temporal dependencies in addition to spectral characteristics \cite{Weninger2015}. Namely, long short-term memory (LSTM) \cite{Hochreiter1997} and gated recurrent unit (GRU) \cite{Cho2014} layers have been incorporated into CNNs \cite{tan_wang_2018, zhao_zarar_tashev2018}. A fully convolutional recurrent network (FCRN) \cite{Strake2020} has been introduced to combine the strengths of convolutional modelling even throughout the recurrent layers by using convolutional LSTMs (CLSTMs) for feature processing, achieving state-of-the-art performance \cite{Strake2020b}.

Considering the reduction of computational complexity, multiple approaches such as pruning or quantization have been proposed \cite{Tan2021, Fedorov2020}. Earlier models achieved efficiency using hybrid processing methods and coarse features \cite{valinrnnoise} while recent methods employed deep filtering in a multi-stage setup \cite{Schroeter2022}. While other fields have seen neural architecture search employed to find efficient base topologies for subsequent scaling \cite{tan2019efficientnet}, in speech enhancement an important recent advancement in terms of providing an efficient high-performance topology was the Convolutional Recurrent U-net for Speech Enhancement (CRUSE) class of networks \cite{Braun2021}. 
However, the reduction of computational complexity comes with a tradeoff in terms of model performance. Additionally, the huge parameter counts, with the smaller CRUSE versions {being} even in the range of the original FCRN ($5.2$ M) or higher, remain a problem for memory-constrained applications.

Our work builds upon the idea of the FCRN and improves the efficiency by reducing the number of filters and the kernel size. This allows to increase the network's depth, but requires to introduce learnable skip connections and an only linear increase (decrease) of filter numbers in the encoder (decoder), thereby creating a smaller network that retains most of its performance. Besides the efficient \texttt{FCRN15}, our core contribution in this paper is the efficient convolutional recurrent neural network \mbox{(EffCRN)} topology which takes the design principles "deeper and wider with smaller kernels" a step further. Additionally, we reduce zero-padding of layer inputs and regain good quality by allowing for non-convolutional layers in the now sparse bottleneck. We compare the FCRN and EffCRN variants to the strongest CRUSE networks.

The paper is structured as follows. Section \ref{sec:ModelTopos} introduces the evaluation framework and the network topologies used in our experiments. Experimental setup, datasets, and training parameters are detailed in Section \ref{sec:Experiments_and_Discussion}, followed by a discussion of our experimental results. We conclude in Section \ref{sec:Conclusion}. 

\section{Model Topologies}
\label{sec:ModelTopos}
\subsection{Evaluation Framework}

Our models operate in the STFT domain. The spectral coefficients of clean speech $S_\ell(k)$ and noise $D_\ell(k)$ yield noisy speech $Y_\ell(k) = S_\ell(k) + D_\ell(k).$
While $\ell$ denotes the frame index, $k \in \mathcal{K} = \{0,1,\dots \allowbreak K \! - \! 1\}$ is the frequency bin index of a $K$-point DFT. The models predict a spectral mask $G_\ell(k) \in \mathbb{C}$ which is the complex-valued representation of the real-valued network output tensor $\mathbf{G}_\ell(k) \in \mathbb{R}^2$. The clean speech estimate is then $\hat{S}_\ell(k) = G_\ell(k) \cdot Y_\ell(k)$.

\begin{figure}[t]
	\centering
	\centerline{
	 \includegraphics[width=1\linewidth]{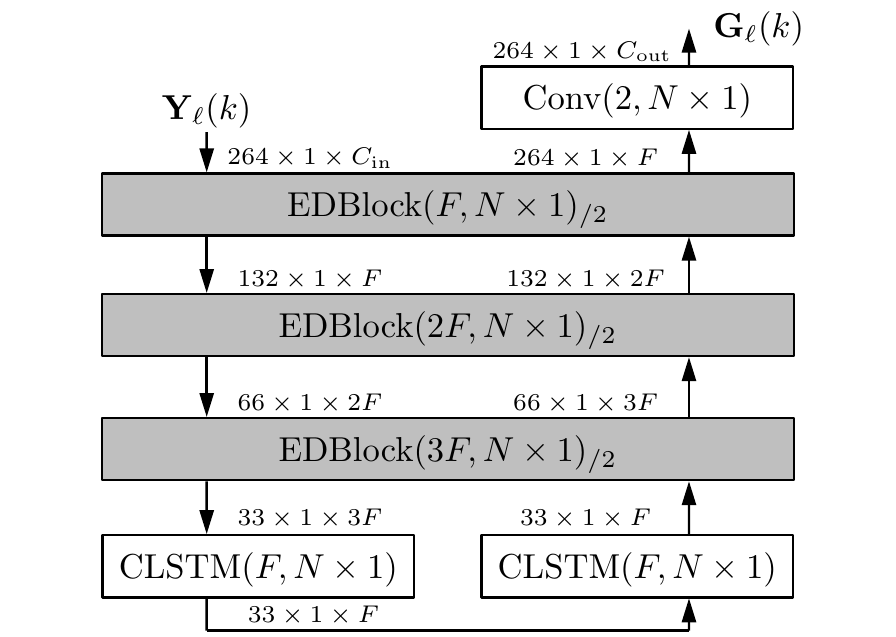}
	 }
	\caption{The \texttt{FCRN15} topology depicted using the EDBlock() detailed in Fig.\ 2}
	\label{fig:fcrn15}
\end{figure}

\subsection{FCRN Variants}

The original FCRN \cite{Strake2020b} 
consists of $10$ convolutional layers, $1$ CLSTM and multiple up-/downsampling layers as well as $2$ skip connections that additively combine features from the encoder and from the decoder. A more efficient version of this network is the \texttt{FCRN15} depicted in Fig.\ 1. 
It performs up- and downsampling with strided convolutions {(as shown to be beneficial in \cite{Strake2020c})} and \textit{increases the network's depth to $15$ layers}. Additionally, it \textit{enhances the additive skip connections with $1\times 1$ depthwise convolutions}. Less and smaller filter-kernels enable its two CLSTMs to be more efficient than the FCRN's single CLSTM. We depict the \texttt{FCRN15} using an encoder-decoder block EDBlock() which combines two convolutions from both encoder and decoder and the learnable skip connections as shown in Fig.\ 2. 
It requires two input signals $\mathbf{x}^{\mathrm{enc/dec}}_\mathrm{in}$ and produces two output signals $\mathbf{x}^{\mathrm{enc/dec}}_\mathrm{out}$. The number of filter kernels is $i\cdot F$, where $i$ is the index of the EDBlock, $N$ and $V=1$ are the size of the {3}D kernel along the frequency and the time axis, respectively, and $S=2$ denotes the stride. {All employed non-depthwise convolutions possess kernels with a third axis covering all available feature maps.} The size of the compressed frequency axis is $M'$, while $C_{\mathrm{in}}^{\mathrm{enc/dec}}$ describes the number of channels of each input.
Convolutional layers are denoted by $\mathrm{Conv}(i\cdot F, N \times V)_{/S}${, with $S$ being an optional stride. $\mathrm{DeConv}(i\cdot F, N \times V)_S$ refers to a transposed convolution with the same parameter set.}
Layer outputs have a size \textit{feature axis $\times$ time axis $\times$ feature maps}. {In comparison to the well known CRUSE topology we only downsample every other convolutional layer and we increase the number of filters linearly instead of exponentially with depth.}

All models take an input $M \times 1 \times C_{\mathrm{in}}$ with $M=K/2+1+P$ and $P$ being the number of zeros padded to the non-redundant bins of the spectrum. 
We use the LeakyReLU activation {\cite{Maas2013}} for all but the last convolutional layer, which uses linear activation. Instead of $G_\ell(k)$, the bounded network output 

\begin{equation}
\label{eq_bounding}
G'_\ell(k) = \mathrm{tanh}(|G_\ell(k)|) \cdot \frac{G_\ell(k)}{|G_\ell(k)|}
\end{equation}

\noindent with constrained magnitude $| G'_\ell(k)| \in [0,1]$ is then used for masking by $\hat{S}_\ell(k) = G'_\ell(k) \cdot Y_\ell(k)$ \cite{Strake2020b}.
CLSTM activations are tanh and sigmoid as published in \cite{Shi2015}.

\begin{figure}[t]
	\centering
	\centerline{
	\includegraphics[width=1\linewidth]{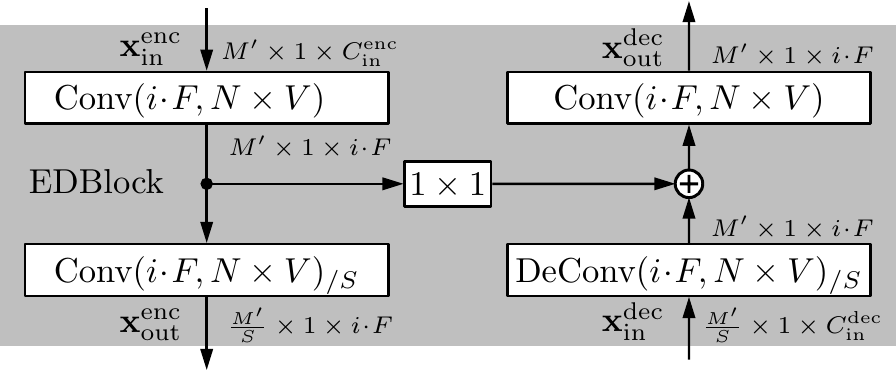}	
	}
	\caption{Details of the $i$-th $\mathrm{EDBlock}(i \cdot F, N \times V)_{/S}$, consisting of four convolutional layers, providing  downsampling on the encoder side and upsampling on the decoder side.}
	\label{fig:quadconv}
\end{figure}

\subsection{The New EffCRN Topologies}

\begin{figure}[t]
	\centering
	\centerline{
\includegraphics[width=1\linewidth]{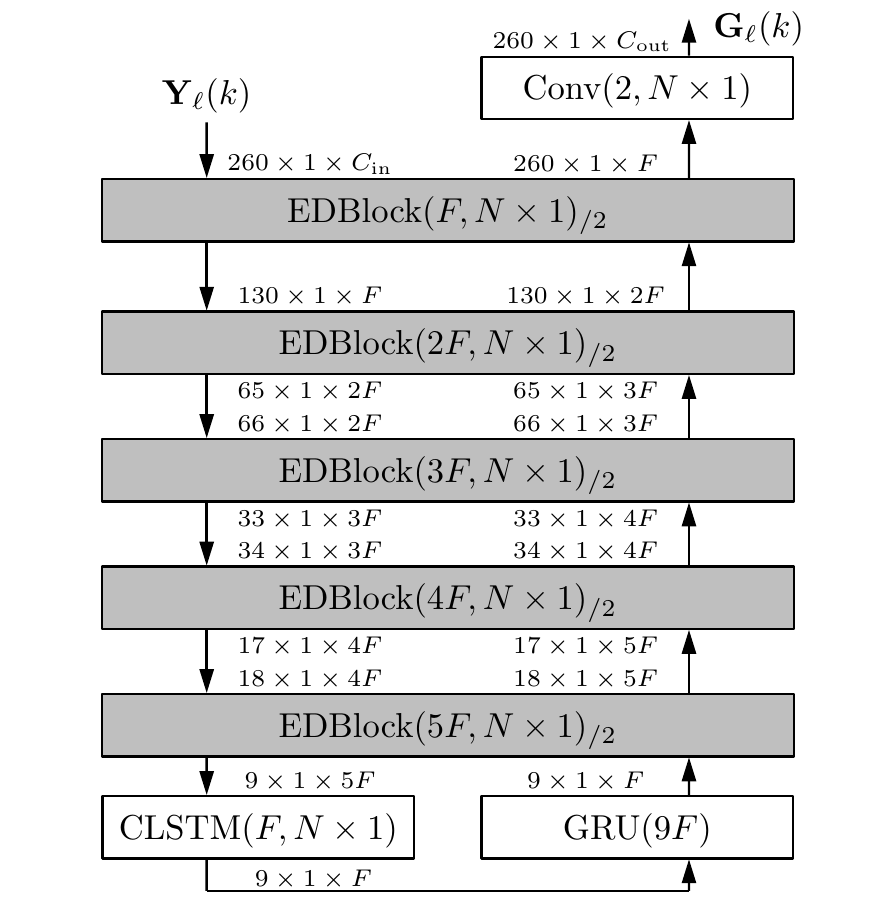}	
	}
	\caption{The \texttt{EffCRN23}/\texttt{EffCRN23lite} topology using the EDBlock detailed in Fig.\ 1. Mismatching dimensions are matched by zero-padding (encoder) or removal (decoder).}
	\label{fig:clg23}
\end{figure}

The new \texttt{EffCRN23} topology can also be depicted using the EDBlocks as shown in \autoref{fig:clg23} with an increased total \underline{d}epth of now $23$ layers (change {\circled{D}}). Compared to the \texttt{FCRN15}, the self-evident approach for the \texttt{EffCRN23} is to initially use fewer and overall significantly smaller \underline{f}ilter kernels (change {\circled{F}}). 
{However, contrary to the CRUSE approach, we do not decrease network depth to obtain the smaller network. Instead we increase network depth, thereby allowing the network to compensate for the drastic impact of reduced filter count and size on it's capacity. The increased depth allows the network to find its own powerful feature representation while maintaining overall computing costs low. This is in line with literature which states that an optimal ratio of depth to width exists for a given network \cite{tan2019efficientnet}. Our network performs further downsampling along the frequency axis and achieves a significantly smaller feature representation at the bottleneck between encoder and decoder, where recurrent modelling takes place. Notably we use the first CLSTM to sharply reduce the number of filters as well for efficient processing.} 
\textit{However, {to keep output signal quality high, the resulting smaller representation not only allows but requires non-convolutional recurrent bottleneck processing}, replacing the second \underline{C}LSTM layer (change {\circled{C}}) with an efficient non-convolutional \underline{G}RU layer (change {\circled{G}}) without an excessive increase of model parameters.}

Additionally, we reduce computations by optimizing zero-\underline{p}adding and apply it only when necessary (change {\textcircled{\raisebox{-0.5pt}{\scriptsize P}}}). Single entries are padded to even input sizes in the encoder directly before EDBlocks and the extra entries are removed at the respective location in the decoder as shown in Fig.\ 3. 
The \texttt{EffCRN23lite} is an even smaller version of the same topology featuring fewer filters. Activations and output bounding (\ref{eq_bounding}) are identical to the FCRN variants.

{Please note that the changes from \texttt{FCRN15} to \texttt{EffCRN23} can be concisely stated by the introduced shorthand as: $\texttt{EffCRN23} = \texttt{FCRN15} \circledp{D} \circledp{P} \circledp{F}  \circledm{C} \circledp{G}$}.

\subsection{CRUSE Network Baselines}

We compare our work against the CRUSE networks \cite{Braun2021}, which have been designed to be efficient. We choose the two most powerful architectures from \cite{Braun2021}, namely CRUSE5\_256-\_2xLSTM1 (\texttt{CRUSE5}) and CRUSE4\_\-128\_1xGRU4\_convskip (\texttt{CRUSE4}) for re-imple\-men\-ta\-tion and reference. They feature a symmetrical encoder-decoder structure consisting entirely of convolutional blocks using a $(3\times 2)$ kernel, whereas our models process single frames $(V=1)$. In the bottleneck they feature either $2$ sequential LSTMs or $4$ parallel GRU layers. The GRUs were fed by flattening, then splitting the input data to those layers.
Since we focus on a network comparison, we employ the CRUSE topologies detailed in \cite{Braun2021}, keeping layers and activations unchanged. Note that applying a linear output layer and a bounding (\ref{eq_bounding}) as in the FCRN/EffCRN variants, performance changes negligibly. We use the same input features as for our networks and pad them to fulfill divisibility constraints imposed by repeated downsampling. 

\section{Experiments and Discussion}
\label{sec:Experiments_and_Discussion}

\subsection{Datasets and Parameter Settings}

Experiments are performed on WSJ0 speech data \cite{Garofalo2007} mixed with noise from the DEMAND \cite{Thiemann2013} and QUT \cite{Dean2010} datasets for training and development, whereas unseen noise data is taken from the ETSI dataset \cite{ETSI2008} for the test set. SNR conditions are $0$, $5$ and $10$ dB at an active speech level of $-26$ 
dBov before mixing \cite{ITU-P56}.
We employ $105$ hours of training data, $10$ hours of validation data, and 1 hour of test data, with disjoint speakers between any of the three datasets. 

All audio data is sampled at $16$\,kHz. Framing is performed with a square-root Hann window / frame length of $512$ samples (equals DFT size $K$) and a $50\%$ frameshift, allowing \#FLOPS/frame to be compared across all networks. Input and output of all networks are real and imaginary parts of spectrum and gain, respectively ($C_{\mathrm{in}}\! = C_\mathrm{out} \!= 2$ channels). 

While the baseline FCRN uses $F\!=88$, $N\!=24$ and $M\!=260$, for the \texttt{FCRN15}, we choose $F\!=32$ and $N\!=12$. The input size $M\!= 264$ enables threefold downsampling by a factor of $2$. For the \texttt{EffCRN23} topology we set the number of filters to $F\!=27$ and the kernel size $N\!=4$. Due to in-network padding we can choose $M\!=260$. The \texttt{EffCRN23lite} uses $F\!=17$ for an even smaller network and is otherwise identical to the \texttt{EffCRN23}. 

\subsection{Network Training}

The models are trained using backpropagation-through-time
with a sequence (utterance) length of $|\mathcal{L}_u|=100$ frames, corresponding to $1.6$ seconds, and a minibatch size of $|\mathcal{B}|= 16$. We use the Adam optimizer with standard parameter settings as given in \cite{Kingma2015}. The learning rate is set to $10^{-4}$, which is dynamically reduced by a factor of $0.6$ upon $4$ consecutive epochs without improved validation loss. Training is stopped after the learning rate falls below $10^{-6}$, or after $10$ consecutive epochs without improved validation loss, or upon completing $70$ epochs. We train using a fixed seed that we do not optimize for. Training is performed on a single \texttt{Nvidia GTX 1080 Ti}.
As optimization target we use the state of the art loss function by Braun and Tashev \cite{Braun2020b}

\begin{align}
\label{braun_loss}
J &= \frac{1}{|\mathcal{B}|} \sum_{u \in \mathcal{B}} \frac{1}{|\mathcal{L}_{u}|} \sum_{\ell \in \mathcal{L}_{u}} \big( \frac{1\!-\!\alpha }{|\mathcal{K}|} \sum_{k\in \mathcal{K}} \big||\hat{S}_\ell(k)|^c - |{S}_\ell(k)|^c\big|^2 \nonumber 
\\&+  \frac{\alpha}{|\mathcal{K}|} \sum_{k \in \mathcal{K}} \big||\hat{S}_\ell(k)|^c e^{j \varphi_{{\hat{\mathrm{s}},\ell}}(k)} - |{S}_\ell(k)|^c e^{j\varphi_{\mathrm{s},\ell}(k)} \big|^2 \big),  
\end{align}

\noindent with a compression factor $c=0.3$ and $\alpha = 0.3$ being a weighting between complex and magnitude contribution. The set of utterance indices per minibatch is denoted as $\mathcal{B}$, 
while $\mathcal{L}_u$ denotes the set of frame indices in utterance $u$. 
The phase of the enhanced and clean signal are denoted as $\varphi_{{\hat{\mathrm{s}},\ell}}(k)$ and $\varphi_{{\mathrm{s}},\ell}(k)$, respectively.

\subsection{Metrics, Results, and Discussion}

For the instrumental evaluation we use perceptual evaluation of speech quality (PESQ) \cite{ITU-P862.2}, $\Delta \mathrm{SNR} = \mathrm{SNR}_\mathrm{out} - \mathrm{SNR}_\mathrm{in}$ and DNSMOS \cite{Reddy2021}. Signal levels for SNR calculation are determined following ITU-P.56 \cite{ITU-P56}. Parameter count and \#FLOPs/frame are reported based on our implementation in \texttt{Tensorflow 2.7} \cite{abadi2016tensorflow}. The evaluation results are averaged over all unseen noise types and all SNRs of the test set and presented in \autoref{tab:results}.

The baseline FCRN \cite{Strake2020b} shows the best performance in all metrics although at a high computational complexity of $1.5$ GFLOPs/frame. The efficient CRUSE networks computationally operate at a significantly lower \#FLOPs/frame, but in our setup require even more parameters than the FCRN. The \texttt{FCRN15} scores close to the baseline in all reported measures, being just $0.06$ PESQ points and $0.19$ dB in $\Delta$SNR below, but with only $16.8\%$ of parameters and $8.2\%$ of computational complexity. The \texttt{EffCRN23} has a slightly higher number of parameters but still stays below $1$ M and reduces computational complexity even further: With $41$ MFLOPs/frame the \texttt{EffCRN23} requires only $3\%$ of computational complexity of the original FCRN at the price of $0.11$ PESQ points and $0.38$ dB $\Delta$SNR. Downscaling the architecture to the \texttt{EffCRN23lite} reduces computational complexity to $1.1\%$ with still moderate losses in PESQ (0.15) and $\Delta$SNR ($0.31$ dB). DNSMOS results are close, with the \texttt{EffCRN23} being the best among the efficient models.

Figs.\ 4 and 5 show an overview of all models' performance in terms of $\Delta$SNR ($\circ$) and PESQ ($\ast$) over \#FLOPS/frame and parameter count {while the respective DNSMOS values can be seen in \autoref{tab:results}.} In the \textit{efficient} network regime, the \texttt{FCRN15} excels over the \texttt{CRUSE5} in all reported metrics PESQ, DNSMOS, $\Delta$SNR, with lower parameter count (-98\%) and \#FLOPS/frame (-32\%). In the \textit{highly efficient} network regime, the \texttt{EffCRN23lite} excels over the \texttt{CRUSE4} in all reported metrics (-94\% parameters and -20\% \#FLOPS/frame), showing that our FCRN/EffCRN family of networks provides best-in-class network topologies for speech enhancement. 

\begin{figure}
	\centering
	\centerline{
	\includegraphics[width=1\linewidth]{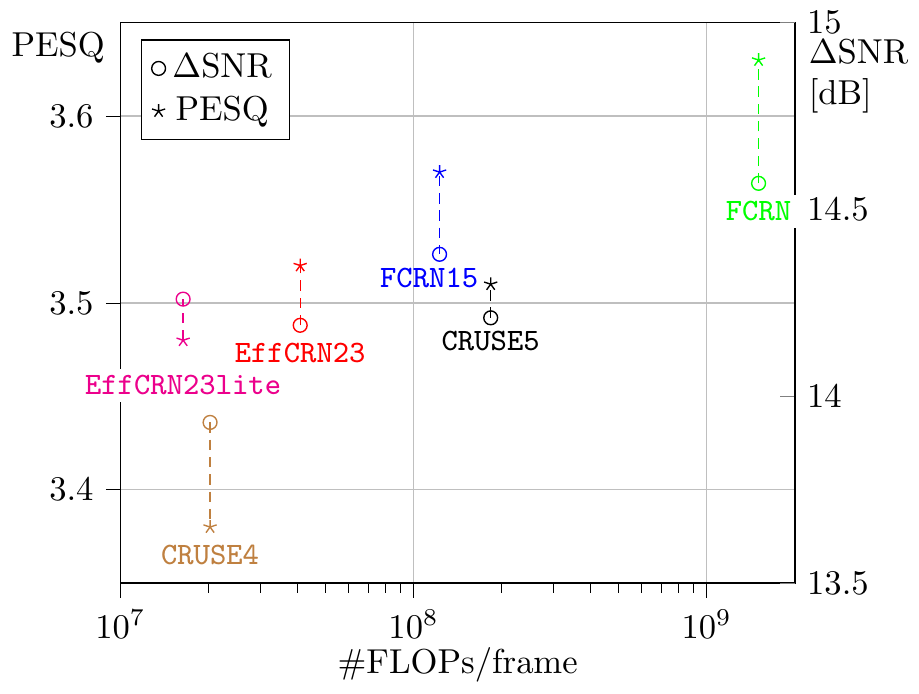}	
	}
	\caption{PESQ and $\Delta$SNR over \#FLOPs/frame. The \texttt{EffCRN23lite} model excels over the \texttt{CRUSE} baselines in PESQ, $\Delta$SNR, and FLOPs/frame. {Data from \autoref{tab:results}.}}
	\label{fig:complexity_plot}
\end{figure}

\begin{table}[t]
	\setlength{\tabcolsep}{.35em}
	\caption{Instrumental evaluation on the test set $\mathcal{D}_{\mathrm{test}}$. Best efficient network results are \textbf{bold}, second best \underline{underlined}. Ablations in bottom part out of competition.}
	\input{tables/TABLE_results_test.tex}
	\label{tab:results}
\end{table}

Furthermore, Table 1 (lower half) and Table 2 show the impact of our intermediate steps (\circledm{C},\circledp{G},\circledp{D},\circledp{P},\circledp{F}) on computational complexity and model performance. Experimenting with the recurrent layers of the \texttt{FCRN15} shows a slight loss of performance in case of both removal of the second CLSTM (\texttt{FCRN15\circledm{C}}) and its replacement with a GRU (\texttt{FCRN15\circledm{C}\circledp{G}}). With $7.2$ M parameters, however, the latter is simply too large. As visible in \autoref{tab:computational_impact}, just increasing network depth alone (\texttt{FCRN15}\circledp{D}) does not reduce, but increase both \#parameters and \#FLOPs/frame, as expected. Also shifting padding from input data towards internal data representations of the network (\texttt{FCRN15}\circledp{D}\circledp{P}), \#FLOPs/frame can be reduced by $9$ M, yet the network remains unacceptably large ($2.8$ M). Beginning our modifications with just decreasing filter numbers and kernel sizes, leads to a very small and efficient network (\texttt{FCRN15}$\circledp{F}$), however, coming with a dramatic performance loss across all measures compared to the \texttt{FCRN15}, see Table \ref{tab:results} again. 
This motivates in a first step to combine the deep network with smaller and fewer filter kernels (\texttt{FCRN15\circledp{F}\circledp{D}\circledp{P}}), which yields a still small ($665$ K) and efficient ($41$ MFLOPS/frame) network. 
To regain performance in a second step, we revisit the initial idea of modifying recurrent processing by replacing the \texttt{FCRN15\circledp{F}\circledp{D}\circledp{P}}'s second CLSTM with a GRU (\texttt{FCRN15}\circledp{F}\circledp{D}\circledp{P}\circledm{C}\circledp{G}) which is in total then identical to the \texttt{EffCRN23}. Due to the small bottleneck feature representation, this network manages to regain {$0.08$} PESQ-points and {$0.09$} DNSMOS-points while maintaining very low complexity and staying below $1$ M parameters.

\section{Conclusions}
\label{sec:Conclusion}

In this paper, we have introduced the \texttt{FCRN15} topology and newly proposed the \texttt{EffCRN23} class of networks for speech enhancement. Significant \hbox{reductions} in model parameter count as well as \#FLOPs/frame compared to the FCRN baseline can be achieved by smaller kernels and an only linear increase (decrease) of the number of filters in the encoder (decoder). This allows then to \hbox{increase} the network depth, but requires learnable skip connections and re-introduction of a \hbox{non-convolutional} (GRU) layer in the sparse bottleneck. The efficient models retain high performance with the \hbox{\texttt{EffCRN23lite}} requiring $7.6\%$ of parameters and $1.1\%$ of FLOPs/frame compared to the baseline topology. We furthermore show that the \texttt{FCRN15} and \texttt{EffCRN23lite} outperform the \texttt{CRUSE5} and \texttt{CRUSE4} networks, respectively, w.r.t.\ PESQ and DNSMOS and $\Delta$SNR, while requiring about 94\% less parameters and about 20\% less \#FLOPs/frame.

\begin{figure}
	\centering
	\centerline{
	\includegraphics[width=1\linewidth]{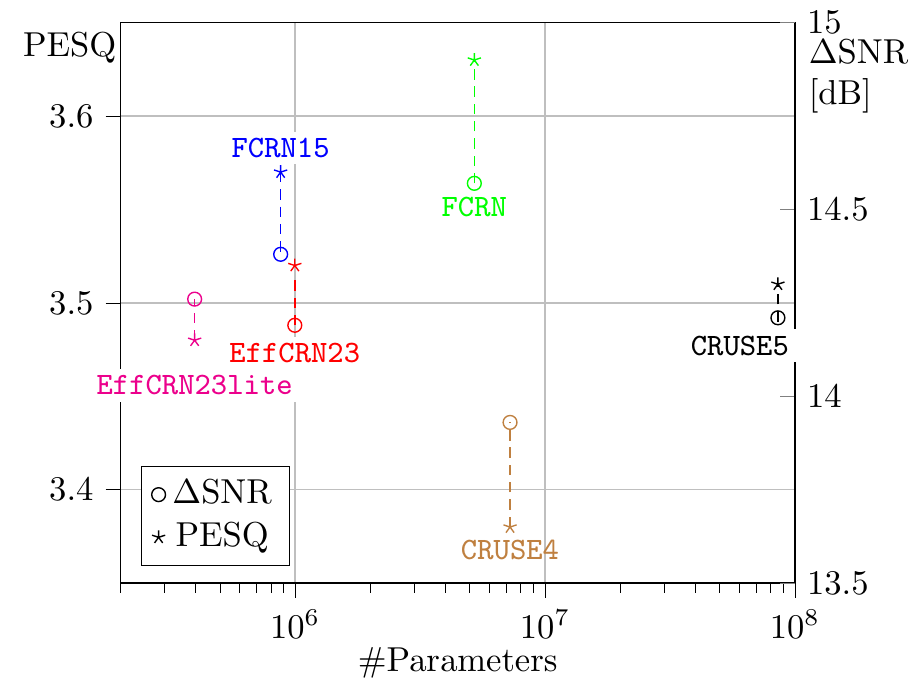}
	}
	\caption{PESQ and $\Delta$SNR over \#parameters. The \texttt{EffCRN23lite} model excels over the \texttt{CRUSE} baselines w.r.t\ PESQ, $\Delta$SNR, and \#parameters. {Data from \autoref{tab:results}.}}
	\label{fig:parameter_plot}
\end{figure}

\begin{table}[t]
	\setlength{\tabcolsep}{.35em}
	\caption{Computational impact of topology modifications.}
	\input{tables/TABLE_computational_impact.tex}
	\label{tab:computational_impact}
\end{table}

\bibliographystyle{IEEEtran}
\bibliography{bibliography/ifn_spaml_bibliography}

\end{document}

%% file: tables/TABLE_results_test.tex
\newcolumntype{R}{>{\raggedleft\arraybackslash}X}
\newcolumntype{C}{>{\center\arraybackslash}X}
\newcommand{\bftab}{\fontseries{b}\selectfont}

\begin{tabularx}{\linewidth}{@{}l r r ccc} 

	\toprule
\multirow{2}{*}{\textbf{Method}} & \multirow{2}{*}{\#par.} & {\#FLOPs/} & \multirow{2}{*}{PESQ} & DNS  & $\Delta\mathrm{SNR}$\\

& & frame & & MOS & [dB] \\
\midrule

 \fontsize{8}{12}\selectfont Noisy & - & - & 2.30 & -  & - \\
 \fontsize{8}{12}\selectfont FCRN \cite{Strake2020b} & 5.2 M & 1500 M & 3.63 & 3.16 & 14.57 \\
  
\midrule

\fontsize{8}{12}\selectfont  \tt{FCRN15} & \underline{875 K} & 123 M &\bftab{3.57} & \underline{3.12} & \bftab{14.38}\\
\fontsize{8}{12}\selectfont  \tt{EffCRN23}  & 997 K & 41 M & 3.52 & \bftab{3.13} & 14.19\\
\fontsize{8}{12}\selectfont  \tt{EffCRN23lite} & \bftab{396 K} & \bftab{16 M} & 3.48  & 3.09 & \underline{14.26}\\
\fontsize{8}{12}\selectfont  \texttt{CRUSE5} \cite{Braun2021} & 85 M & 183 M & \underline{3.53} & 3.11 & 14.20\\
\fontsize{8}{12}\selectfont  \texttt{CRUSE4} \cite{Braun2021} & 7.2 M & \underline{20 M} & 3.38  & 3.03 & 14.06 \\

\midrule

\fontsize{8}{12}\selectfont \tt{FCRN15}$\circledm{C}$ & 777 K & 112 M & 3.54 & 3.12 & 14.30\\
\fontsize{8}{12}\selectfont \tt{FCRN15}$\circledm{C}\circledp{G}$ & 7.4 M & 125 M & 3.51 & 3.13 & 14.17\\
\fontsize{8}{12}\selectfont \tt{FCRN15}$\circledp{F}$ & 209 K & 29 M & {3.41} & {3.05} & {14.14} \\
\fontsize{8}{12}\selectfont \tt{FCRN15}$\circledp{F}\circledp{D}\circledp{P}$ & 665 K & 41 M & {3.44} & {3.04} & {14.07} \\

\bottomrule
\end{tabularx}

%% file: tables/TABLE_computational_impact.tex
\newcolumntype{R}{>{\raggedleft\arraybackslash}X}
\newcolumntype{C}{>{\center\arraybackslash}X}
\newcommand{\bftab}{\fontseries{b}\selectfont}

\begin{tabularx}{\linewidth}{l R R} 
\toprule
\textbf{Method} &\textbf{\#par.} & \textbf{\#FLOPs/frame}\\
\cmidrule(lr){1-3}
\tt{FCRN15} & 875 K & 123 M  \\
\tt{FCRN15}\circledm{C} & 777 K & 112 M  \\
\tt{FCRN15}\circledm{C}\circledp{G} & 7400 K & 125 M  \\
\tt{FCRN15}\circledp{D} & 2800 K & 183 M  \\
\tt{FCRN15}\circledp{D}\circledp{P} & 2800 K & 172 M \\
\tt{FCRN15}\circledp{F} & 209 K & 29 M \\
\tt{FCRN15\circledp{F}\circledp{D}\circledp{P}} & 665 K & 41 M  \\

\bottomrule
\end{tabularx}